\documentclass[letter,traditabstract]{aa}  
\usepackage{graphicx}
\usepackage{txfonts}
\usepackage{natbib}
\usepackage{authblk}
\usepackage{amssymb}
\usepackage{subfig}

\begin{document}
   \title{Observational evidence of quasar feedback quenching star formation
   at high redshift \thanks{Based on data obtained at the VLT through
   the ESO program 077.B-0218(A).}}

   \author{ M. Cano-D\'{\i}az\inst{1}
   \and
          R. Maiolino\inst{1,2}
		  \and
	  A. Marconi\inst{3}
	  \and
	  H. Netzer\inst{4}
	  \and
	  O. Shemmer\inst{5}
	  \and
	  G. Cresci\inst{6}
          }

   \institute{INAF-Osservatorio Astronomico di Roma, via di Frascati 33, 00040 Monteporzio Catone, Italy
         \and
	Cavendish Laboratory, University of Cambridge, 19 J. J. Thomson Ave., Cambridge CB3 0HE, UK
         \and
             Dipartimento di Fisica e Astronomia, Universit\`a degli Studi di Firenze, Largo E. Fermi 2, 50125 Firenze, Italy
         \and
School of Physics and Astronomy and the Wise Observatory
Tel-Aviv University, Tel-Aviv 69978, Israel		 
         \and
Department of Physics, University of North Texas, Denton, TX 76203, USA
         \and
	INAF-Osservatorio Astrofisico di Arcetri, Largo E. Fermi 5, 50125 Firenze, Italy
             }

   \date{Received ; accepted }

  \abstract  
   {Most galaxy evolutionary models require quasar feedback to regulate star formation
   in their host galaxies. In particular, at high redshift, models expect that
   feedback associated with quasar-driven outflows 
   is so efficient that the gas in the host galaxy is largely swept
   away or heated up, hence suppressing star formation in massive galaxies.
   We observationally investigate this phenomenon by using VLT-SINFONI integral field
   spectroscopy of the luminous quasar 2QZJ002830.4-281706 at z=2.4.
   The spectra sample the optical emission lines redshifted into the near-IR.
   The [OIII]$\lambda$5007
   emission-line kinematics map reveals a massive outflow on scales of several kpc.
   The detection of narrow H$\alpha$ emission reveals star formation in the quasar
   host galaxy, with $\rm SFR\sim 100~M_{\odot}~yr^{-1}$.
   However, the star formation is not distributed uniformly, but is strongly suppressed in
   the region with the highest outflow velocity and highest velocity dispersion.
  This result indicates that star formation in this region is strongly quenched by the quasar
  outflow, which is cleaning the galaxy disk of its molecular gas. This is one of the
  first direct observational proofs of quasar feedback quenching the star formation at high
  redshift.
   }
   \keywords{Galaxies: formation -- Galaxies: high-redshift -- Galaxies: evolution
   -- quasars: emission lines}

	\authorrunning{Cano-D\'{\i}az et al.}
	\titlerunning{Quasar feedback quenching star formation at high redshift}
   \maketitle
%
%________________________________________________________________

\section{Introduction}

   Most of the recent galaxy formation models invoke energetic
   outflows as a way to regulate the evolution of galaxies throughout the cosmic
   epochs \citep{Silk,Bower,Springel}. In particular,
   quasars are expected to drive powerful outflows that eventually
   expel most of the
   gas in their host galaxies, thereby quenching both star formation and 
   further black hole accretion \citep[e.g. ]{Granato,DiMatteo,Menci,Bower,Hopkins,King}.
   According to those models, these
   quasar driven outflows are required to prevent massive galaxies from overgrowing,
   hence explaining the shortage of very massive galaxies in the local
   universe, and are responsible for the red color and gas poor properties
   of local elliptical galaxies.

   Massive, large-scale outflows have been
   detected in the hosts of local quasars \citep[e.g. ][]{Feruglio,Fischer,Sturm,Rupke}.
   However, models expect that most of the quasar feedback action occurs at
   high redshift, when quasars reach their peak activity (z$\sim$2) and
   when star formation in the most massive galaxies is observed to
   decline.
   %\citep[e.g. ][]{PerezG,Serjeant}.
   Evidence of outflows in luminous
   quasars has been detected up to very high redshift \citep[e.g.][]{Allen,Maiolino,Alexander}.
  Indications that the strength of these AGN driven outflows anticorrelates with the
 starburst contribution to the infrared luminosity has been obtained in reddened quasars by
 \cite{farrah11}.
   However, direct observational evidence that high-z
   quasar driven outflows quench star formation
   in their host galaxies is still missing.

Here we present VLT-SINFONI near-IR integral field spectra
of the quasar 2QZJ002830.4-281706 (hereafter 2QZ0028-28). This object, at $z = 2.401$,
was taken from the sample of strong [OIII]$\lambda$5007 emitters discovered by \cite{Shemmer},
and it is one of
the most luminous quasars known.
 In this letter we show that the spatially resolved kinematics of the [OIII]$\lambda$5007 line
clearly reveals a prominent outflow on scales of several kpc.
Even more interesting, the star formation traced by narrow H$\alpha$ emission is
suppressed in the region characterized by the strongest outflow.
We suggest that this is the first 
observational evidence of feedback associated with a quasar-driven outflow that
quenches star formation at high redshift, or one of the first.

\begin{figure}[!]
%	\vspace{-11pt}
	\centering
	\includegraphics[height=0.95\linewidth,angle=90]{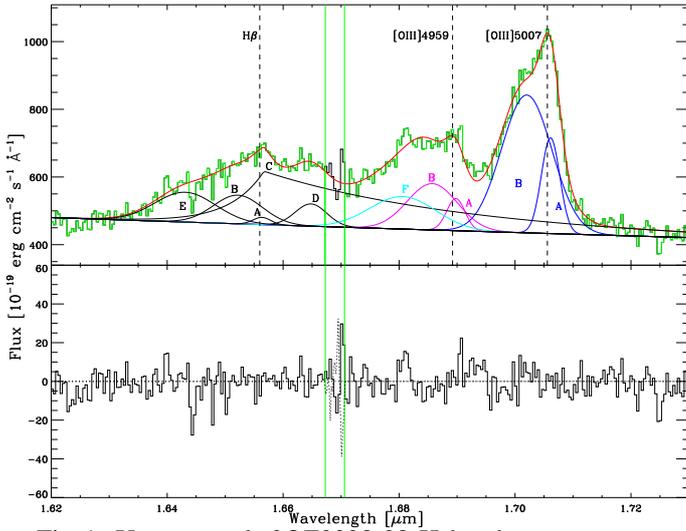}
	\vspace{0.05cm}
	\caption{Upper panel: 2QZ0028-28 H band spectrum extracted from the central 0.5 arcsec, along with the
	various components used for the fit (see Appendix~\ref{app1} for details). Vertical dashed lines indicate
	the rest frame wavelength of each line, by taking the [OIII]$\lambda$5007 line peak for reference.
	Lower panel: Residuals of the fit. The green vertical lines enclose a the spectral zone
	affected by strong
	sky line residuals.}
	\label{hband}
%	\vspace{-22pt}
\end{figure}

\begin{figure*}[!]
  %\centering
  \includegraphics[height=0.4\textwidth,angle=90]{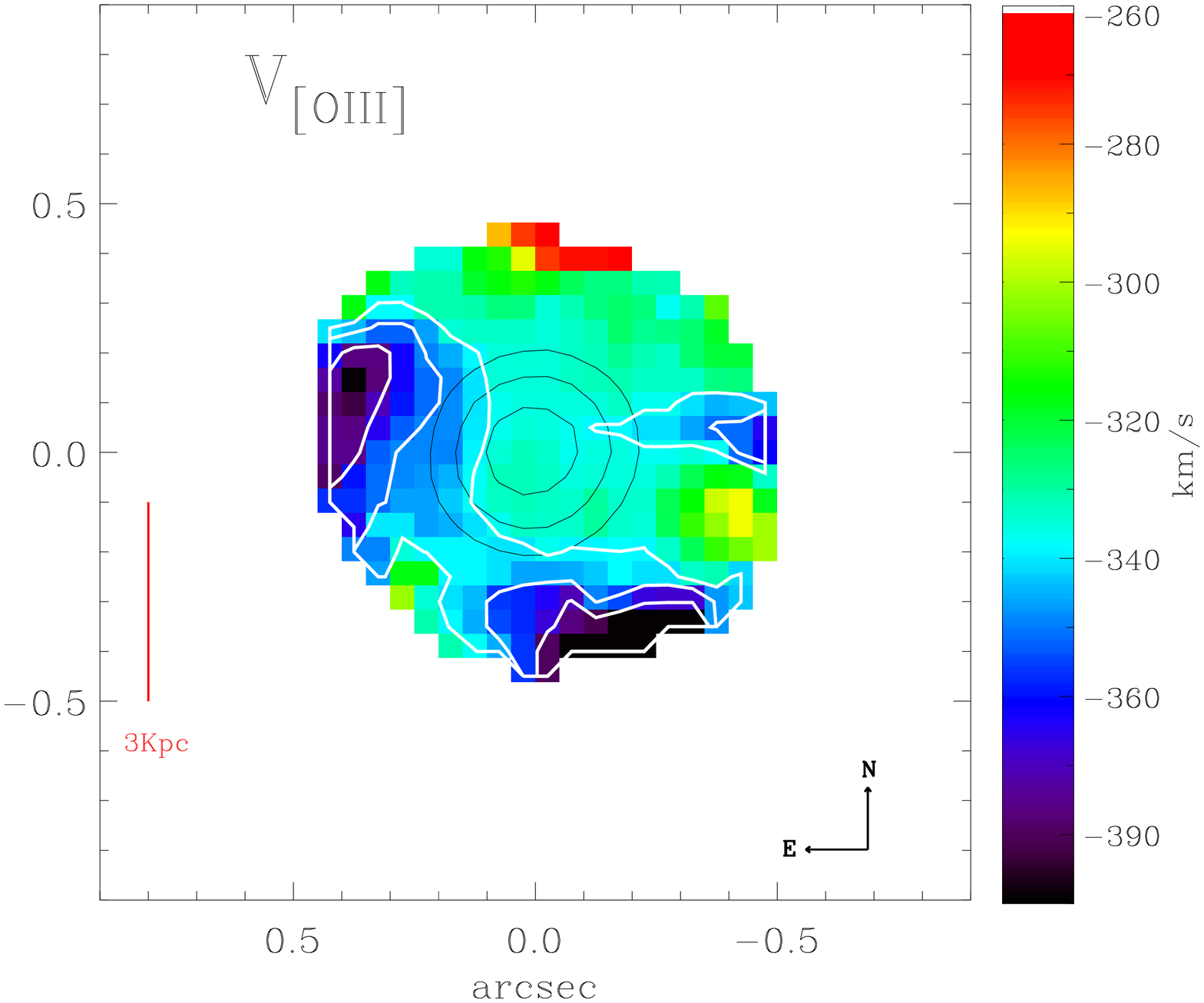}                
  \hspace{1.7cm}
  \includegraphics[height=0.4\textwidth,angle=90]{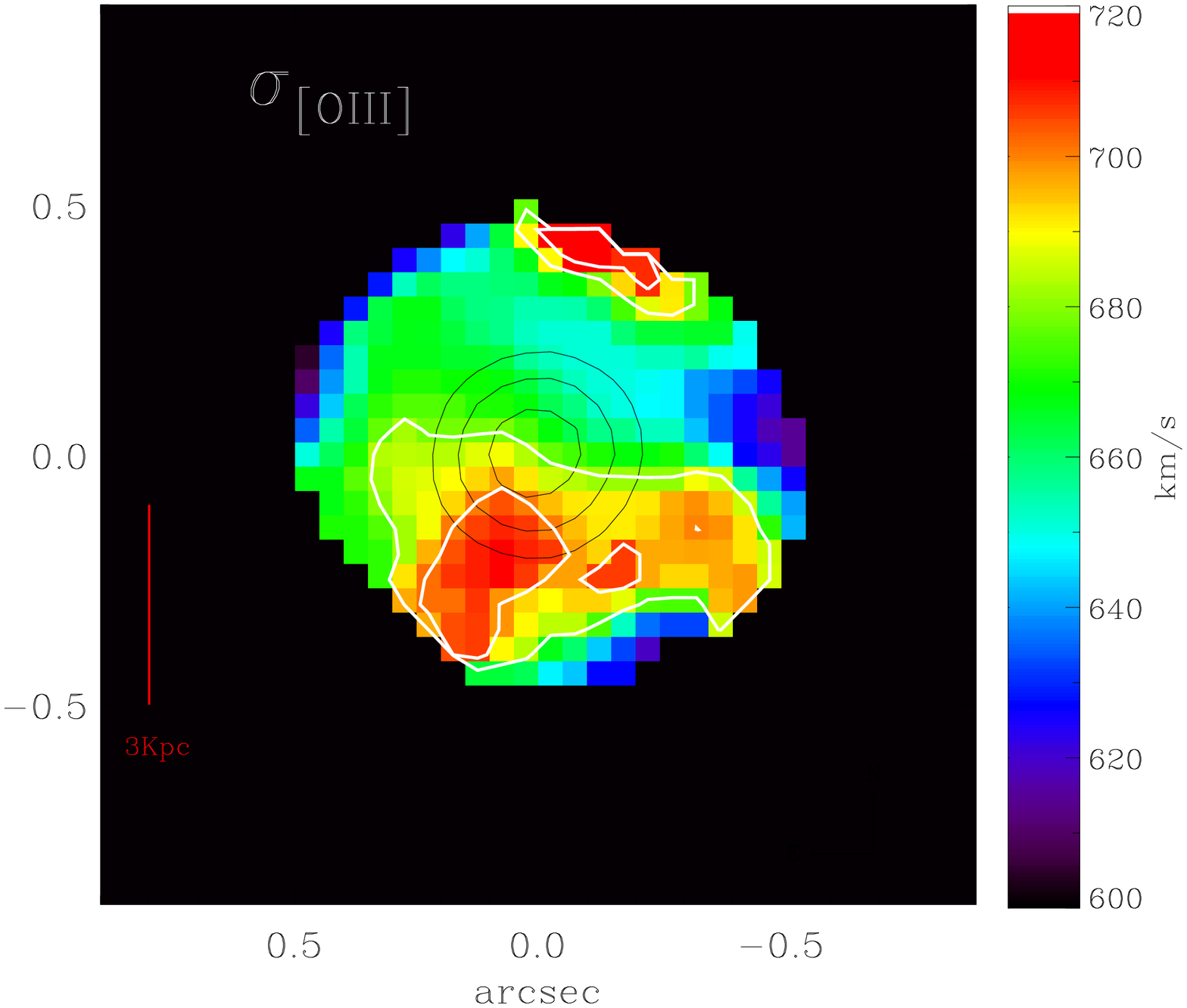}
  \caption{Left panel: Velocity field (first moment map) of the [OIII]$\lambda$5007 line,
  showing the prominent excess of
  blueshifted gas with a bow-like morphology SE of the nucleus. White contours are at 330,
  360, and 390 km/s.
   Right panel: Velocity dispersion (second moment map) of the [OIII]$\lambda$5007 line,
   showing the excess of dispersion in the
  SE region. White contours are at 680 and 700 km/s.
In both maps the black contours trace the continuum.}
  \label{moments}
\end{figure*}

\section{Observations and data reduction}\label{obs}

We used the near-IR integral field spectrometer SINFONI at the VLT to
observe 2QZ0028-28 in the H and K bands, where H$\beta$+[OIII]$\lambda$$\lambda$4959,5007 and
H$\alpha$+[NII]6548,6584 are redshifted, respectively, with the high-resolution gratings
(delivering R=3000 and R=4000, respectively).
On-source exposure times were 1200s in the H band and 2400s in the K band (interleaved with
sky observations). The observations were taken under excellent seeing
conditions: 0.4$''$ (measured from the broad lines,
as discussed in the following).
To properly sample this excellent seeing, we used the camera that delivers
a pixel scale of $0''.1\times0''.05$, providing a field of view of $\rm 3''\times 3''$.

Data was reduced with the ESO-SINFONI pipeline.
The pipeline subtracts the background, performs the flat-fielding, 
spectrally calibrates each individual slice, and then reconstructs the
cube. The pipeline delivers cubes where the spatial pixels are resampled to
$\rm 0''.05\times0''.05$.
%Spectral pixels heavily affected by strong
%sky emission lines were however removed by the analysis.

\section{Data analysis and results}\label{analisis}

The final H and K band data cubes were analyzed by fitting
each spatial spectrum separately in the field of view.
 Details of the fitting procedure for the spectra are given in Appendix \ref{app1}.

\subsection{[OIII]+H$\beta$ analysis: evidence for a powerful outflow}

The H-band spectrum (Fig.~\ref{hband}) shows a clear broad H$\beta$,
associated with the broad line region (BLR),
and a prominent
[OIII] doublet, primarily associated with the quasar narrow line region (NLR).
The [OIII]$\lambda$5007 line shows a clear asymmetric profile with a prominent
blueshifted wing. This is generally a signature of outflows. Indeed,
the lack of a corresponding
redshifted wing is generally due to dust (in the host galaxy) absorbing the wind component
on the opposite side of the galaxy disk, relative to our line of sight.

The [OIII] line is well fitted with two Gaussians as seen in Fig.~\ref{hband}. However,
since the two [OIII] components are heavily blended, their fits are not independent.
To avoid the possible ambiguities associated with
the two components possibly being correlated, we investigate the [OIII] line
kinematics by mapping the first and second moments of the global [OIII] profile
(i.e. average velocity shift and dispersion). These two maps are shown in
Fig. \ref{moments}. Regions with S/N$<$3 were masked out.
As a zero velocity reference we have taken
the peak of the [OIII] line in the central region (Fig.~\ref{hband}).
The black contours
indicate the location of the continuum. As shown by Fig.~\ref{moments}(left),
the average velocity is strongly blueshifted with respect to the line peak, by a few hundred
km/s, as a consequence of the strong blue asymmetry of the line.
The velocity is strongly negative
over the whole region where [OIII] is detected, suggesting that,
if the prominent blue wing is due to a wind, then the outflowing ionized
gas is distributed over all of the central $\sim$7~kpc. However, the most interesting
features are the strongly blueshifted regions to the south and to the
east of the nucleus, together making a bow-like morphology, suggestive of the envelope of
a strong conical outflow. However, the observed velocity gradient may also be
associated with rotation of the host galaxy disk (although the velocity field
is certainly not a typical rotation curve). Outflow and galaxy rotation can
be distinguished through the velocity dispersion: outflows are always associated
with high velocity dispersion compared with regular motions in galaxy disks
\citep[e.g.][]{muller11}.
The velocity dispersion map (Fig.~\ref{moments}) shows a large velocity
dispersion over the whole region where [OIII] is clearly detected, suggesting that the
whole central region is characterized by outflows. However, the
velocity dispersion increases significantly in the SE region, where
the strongly blueshifted gas is detected. This correspondence between
blueshift and velocity dispersion strongly
suggests that the blueshift observed in this region is due to an outflow excess
and not to galaxy rotation.

\begin{figure}[h]
\centering
\includegraphics[height=0.95\linewidth,angle=90]{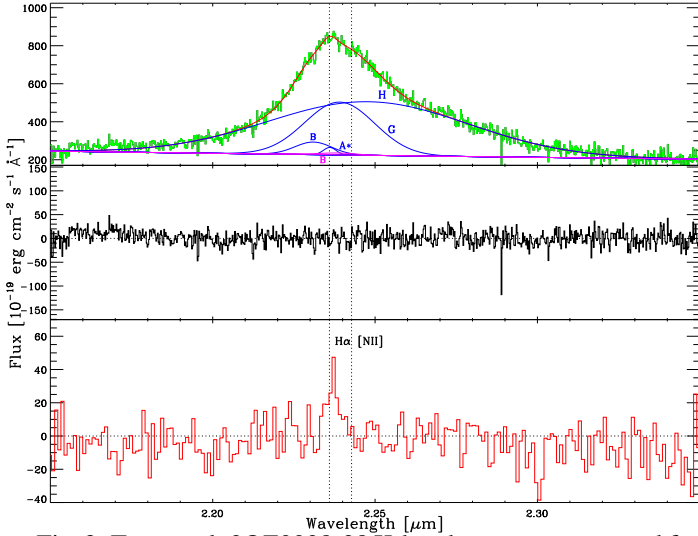}
	\vspace{0.05cm}
\caption{Top panel: 2QZ0028-28 K-band spectrum extracted from the central arcsec,  along with the
	various components used for the fit
	(blue are for H$\alpha$ components, magenta are for [NII]
	lines of component B).
	Vertical dotted lines show the rest-frame wavelength for H$\alpha$
	and [NII]$\lambda$6584, by using the same reference as in Fig.~\ref{hband}. Middle
panel: Residuals of the fit. Bottom panel:
Result of the subtraction between the spectrum extracted from the region with narrow $H\alpha$
emission NW of the nucleus (Fig.~\ref{Ha_fluxmap}) and the spectrum extracted from the region
without narrow $H\alpha$ emission, SE of the nucleus, after scaling the two spectra
to match the intensity of the broad line.
A clear narrow H$\alpha$ component is detected, illustrating that the detection of
this component and its distribution are not artifacts of the spectral fit.} \label{kband}
\end{figure}

\begin{figure*}[!]
 % \centering
  {\includegraphics[height=0.4\textwidth,angle=90]{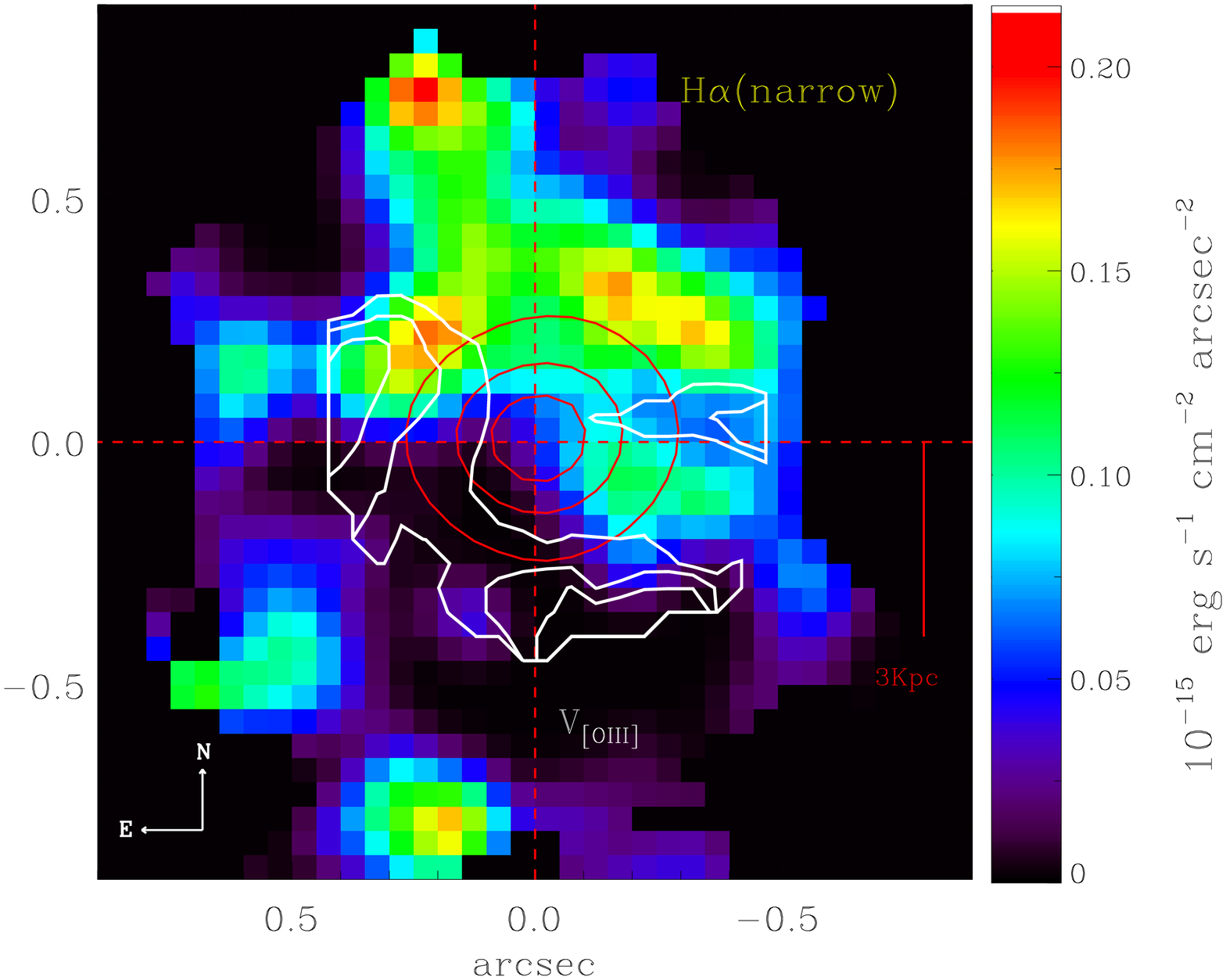}}         
  \hspace{1.7cm}
  {\includegraphics[height=0.4\textwidth,angle=90]{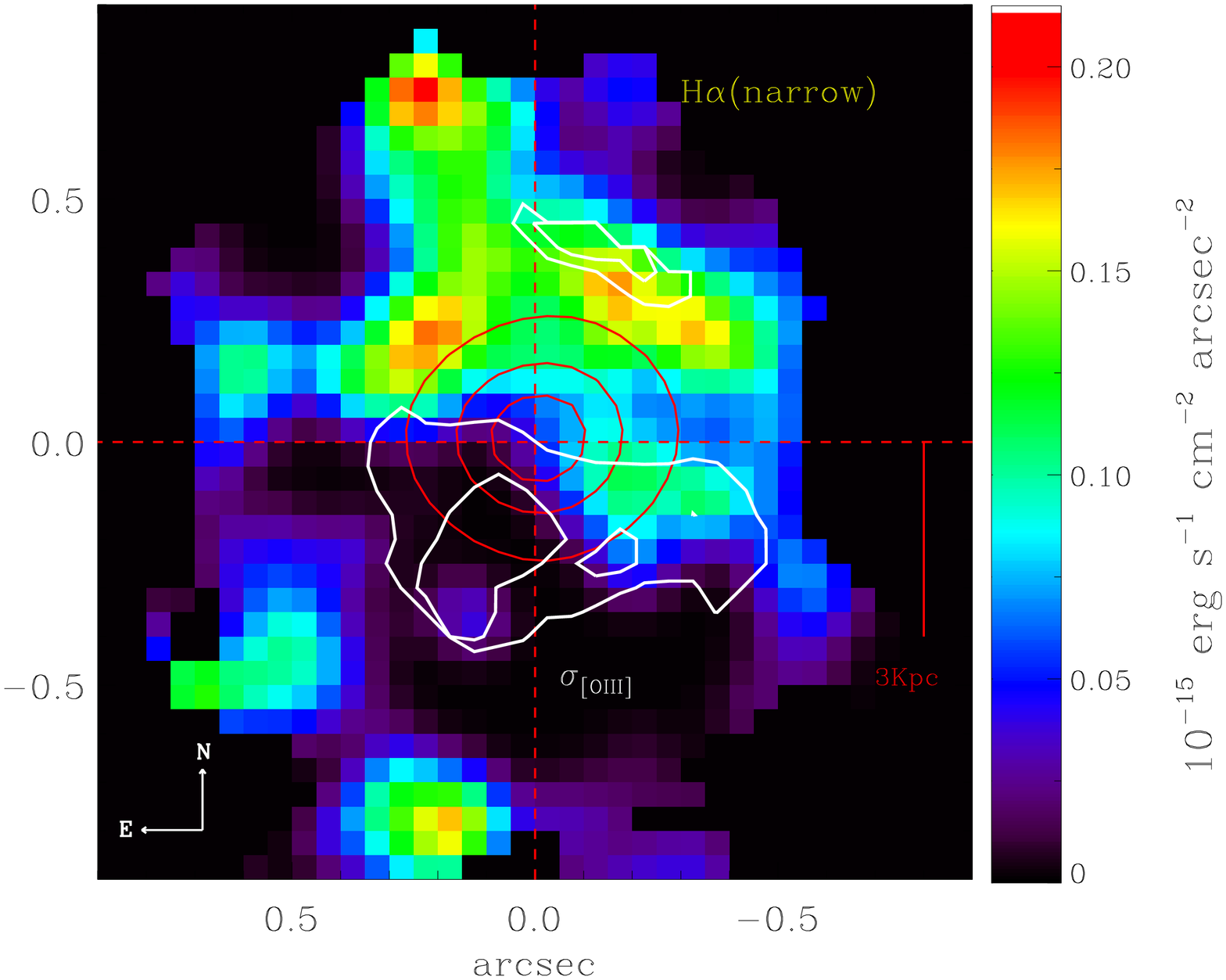}}
  \caption{Map of the narrow component of H$\alpha$ with contours tracing the [OIII] velocity shift
  (left panel) and velocity dispersion (right panel), as in Fig.~\ref{moments}.
  Star formation, traced by H$\alpha$,
  is heavily suppressed in the SE region where the strongest outflow is traced by [OIII]. }
  \label{Ha_fluxmap}
\end{figure*}

While there is a general correspondence between velocity blueshift and dispersion, both
with the highest values in the SE quadrant, a detailed analysis reveals interesting
differences. The peak of the velocity dispersion is located between
the two peaks of velocity blueshift. A possible interpretation is that those areas with highest velocity
dispersion are associated with regions where the outflow is strongly interacting with the gas
in the host galaxy disk, which also lowers the velocity locally.

The outflow may even be intrinsically symmetric, but the opposite (receding) outflow
is likely to be obscured by dust in the host galaxy disk \citep[as observed in local
AGNs,][]{muller11}. It is interesting
to note that the velocity dispersion map shows a strong excess in a tiny region on the NW (opposite
to the main [OIII] outflow), this may be tracing
the opposite outflow coming out of the galactic dusty disk.

The [OIII] line is primarily tracing gas in the NLR
ionized by the QSO. Therefore,
the most likely explanation is that the outflow is driven by the QSO radiation pressure.
A quasar origin of the wind
is also supported by the fact that the outflowing gas reaches velocities in excess
of 1000~km/s (in the [OIII] blue wing), which cannot be explained by models of supernovae-driven
outflows \citep{Thacker}.
The mass of ionized outflowing gas is simply given by (see Appendix~\ref{app2})
\begin{equation}
\rm M_{ion}^{out} = 5.3~10^7 ~\frac{L_{44}([OIII])}{n_{e3}~10^{[O/H]}}~M_{\odot}
\end{equation}
where $\rm L_{44}([OIII])$ is the luminosity of the [OIII]$\lambda$5007 line
tracing the outflow, in units
of $\rm 10^{44}~erg/s$, $\rm n_{e3}$ is the electron density in the outflowing gas, in
units of $\rm 10^3~cm^{-3}$, typical of the NLR, and $\rm 10^{[O/H]}$ is the oxygen abundance is Solar
units.
If we assume that the outflow is traced by the broad component of [OIII],
then $\rm L_{44}([OIII])\sim 2$, implying that the mass of ionized gas involved in the outflow
is $\rm \sim 10^8~M_{\odot}$.

By assuming a simplified
conical (or biconical) outflow 
distributed out to a radius $\rm R_{kpc}$ (in units of kpc), then the 
outflow rate of ionized gas is given by (see Appendix~\ref{app2})

\begin{equation}
\rm \dot{M}_{ion}^{out} = 164~
  \frac{L_{44}([OIII])~v_3}{n_{e3}~10^{[O/H]}~R_{kpc}}~M_{\odot}~yr^{-1}
\end{equation}

where $\rm v_{3}$ is the outflow velocity in units of $\rm 1000~km~s^{-1}$. 
The maximum
outflow velocity inferred from the [OIII] profile is about 2000~km/s, which is probably representative
of the average outflow velocity,
while the lower velocities observed in the [OIII]
line profile are very likely projection effects. By using $\rm R_{kpc}=3$,
we infer an ionized gas outflow rate of about
$\rm 200~M_{\odot}~yr^{-1}$. This is a lower limit of the {\it total} outflow rate, since the ionized
component is probably
a minor fraction of the global outflowing gas mass. If we scale by the same
neutral-to-ionized fraction as in Mrk231 (Rupke, priv. comm.),
the total outflow rate can be up to an order of magnitude higher.

The inferred kinetic power of the ionized component of the outflow is about $\rm 3~10^{44}~erg~s^{-1}$
(Eq.\ref{eq_a9}), or about $\rm \sim 10^{-3}$ times the bolometric luminosity of the AGNs
($\sim 3.8~10^{46}~erg~s^{-1}$,
from the continuum luminosity at 5100\AA , and by assuming a bolometric correction
factor of 7).
However, if the ionized outflow is accompaneid by a neutral/molecular outflow an order of magnitude
more massive, as discussed above, then the kinetic power is also likely to be
an order of magnitude higher, i.e.
about $\sim$1\% of the bolometric luminosity, close to the expectation of quasar feedback models
\citep{lapi05}.

\subsection{H$\alpha$+[NII] analysis: evidence for quenched star formation}
\label{sec_ha}

The $H\alpha$ has a very broad profile (from the BLR), and its overall profile has been fitted mostly with
two broad Gaussians (see Appendix~\ref{app1} for details).
Interestingly, the fit also requires the presence of a weak, but
significant narrow component of H$\alpha$ (A*, FWHM$\sim 600 km/s $).
The absence of [NII] indicates that this narrow
H$\alpha$ emission is mostly due to star formation. 

The map of the narrow H$\alpha$ emission is shown in Fig.\ref{Ha_fluxmap}, which reveals star formation
extending over a few kpc from the nucleus. However, the star formation is
not distributed symmetrically around the nucleus, but primarily towards the N and W.
In particular, the region within a few kpc SE of the nucleus is nearly free of any narrow
H$\alpha$ tracing star formation. To show that the detection of narrow
H$\alpha$ and that the asymmetric
distribution are not artifacts of the fitting procedure, we extracted a spectrum by integrating
over the NW region putatively containing star formation, according to the H$\alpha$ map, and another spectrum
in the SE region devoid of star formation.
After scaling the two spectra so that the intensity of the broad
H$\alpha$ wings (associated with the BLR) are the same, we have obtained the difference of the two
spectra, which is shown in the bottom panel of Fig.~\ref{kband}. The differential spectrum clearly shows
the presence of the narrow H$\alpha$. The line broadened profile (also making the
detection noisier than expected ) is because we are integrating
over different regions of the velocity field. The differential spectrum also confirms that no [NII]
is detected in association with the narrow H$\alpha$, confirming that the latter is tracing
star formation and not the quasar NLR.

We mention that the map of the [OIII] narrow component
(shown in the appendix)
is also characterized by
a similar asymmetry towards the NE, suggesting that some of the [OIII] narrow line is associated
with star formation. However, the asymmetry is less clean than observed for
the H$\alpha$ narrow component, likely
because of NLR contribution to [OIII] and because of the blending with the ``broad'' [OIII] component.

The integrated emission of the narrow H$\alpha$ yields a total star formation rate in the host galaxy
of about $\rm 100~M_{\odot}~yr^{-1}$
\citep[by using the conversion factor given in ][]{Kennicutt},
which is not unusual in high-z quasars \citep[e.g. ][]{Lutz}. However, the most interesting
result is that the star formation is heavily suppressed in the SE region, which is characterized by
the excess of outflow with high-velocity dispersion. In Fig.~\ref{Ha_fluxmap}(left)
the white contours identify
the strongest gas outflow traced by the highly blueshifted [OIII] line, as in Fig.~\ref{moments}-left,
while in Fig.~\ref{Ha_fluxmap}-right the white contours identify the highest velocity dispersion region,
as in Fig.~\ref{moments}-right, which is likely the region where
the strong outflow interacts with the host galaxy disk.
We suggest that the heavy suppression of star formation in the region of strongest quasar-driven
outflow
among the
first direct observational proofs of quasar feedback onto the host galaxy quenching star formation
at high redshift, as predicted by models.

\section{Conclusions}

By using near-IR integral field spectroscopic observations we have revealed a powerful outflow
in the host galaxy of the quasar 2QZ0028-28 at z=2.4.
The outflow was revealed by the velocity field traced by the [OIII]$\lambda$5007 line, redshifted into the
H-band. We estimated that the outflow rate of ionized gas is about $\rm 200~M_{\odot}~yr^{-1}$, which
is, however,
a lower limit of the total gas outflow rate. Both the high
outflow velocity ($\rm > 1000~km/s$)
and the fact that the wind is mostly traced by the [OIII] line (produced primarily in
the NLR) strongly suggest that the outflow is mostly driven by the quasar.
The outflow is not symmetric, the highest velocities
and highest velocity dispersion are found in the region SE of the nucleus.

In the K-band, our data clearly reveal the presence of narrow H$\alpha$ emission tracing star formation
in the host galaxy, on scales of several kpc and with a rate of about $\rm 100~M_{\odot}~yr^{-1}$.
However, star formation is not distributed uniformly in the host galaxy, but is mostly found in the
regions not directly invested by the strong outflow. Instead, star formation is heavily suppressed
in the SE region where the strongest outflow is detected. This observational
result supports models invoking quasar feedback to quench
star formation in massive galaxies at high redshift.

\begin{acknowledgements}
We are grateful to the referee for his/her very useful comments.
MCD is supported by the Marie Curie Initial Training Network ELIXIR under the contract PITN-GA-2008-214227 from the European Commission.
\end{acknowledgements}

\begin{appendix}

\section{Details of the spectral fitting}\label{app1} 

The initial spectral fit is performed on a spectrum extracted from a central
aperture of ten pixels (i.e. 0.5 arcsec) in the H band and 20 pixels (i.e. 1.0 arcsec) in the
K band, which guarantee high S/N.

The emission lines are fitted with multiple Gaussians and, in the case of the broad lines, by also using
power-law profiles, as in \cite{Nagao}. The continuum is fitted with a single power law, which represents the
emission of the QSO accretion disk (plus possibly some minor contribution from the host galaxy
stellar continuum). Starting
from this initial fit in the central region,
we fitted the spectra individually at all spatial pixels by leaving most of the parameters
free, except for the velocity, dispersion, and relative intensity
of the components describing the broad lines (H$\alpha$ and H$\beta$).
Indeed, since the BLR is unresolved, the shape and shift of the broad line profile
(components C+D+E in the case of H$\beta$)
must be constant over the field of view, and there the global intensity
variation only reflects the seeing PSF.
We tried to also include an Fe~II template in the spectral fitting \citep[as in ][]{Netzer};
however, this is always set to zero by the fitting procedure, confirming the lack of significant FeII emission
inferred by the visual inspection of the spectrum. This result contrasts with the measurement of
\cite{Netzer}, who obtain a flux of the Fe~II emission that is
about 0.37 times the H$\beta$ emission, and we ascribe the
discrepancy to the much lower signal-to-noise in the latter spectrum.

\begin{figure*}[!]
 % \centering
  {\includegraphics[height=0.4\textwidth,angle=90]{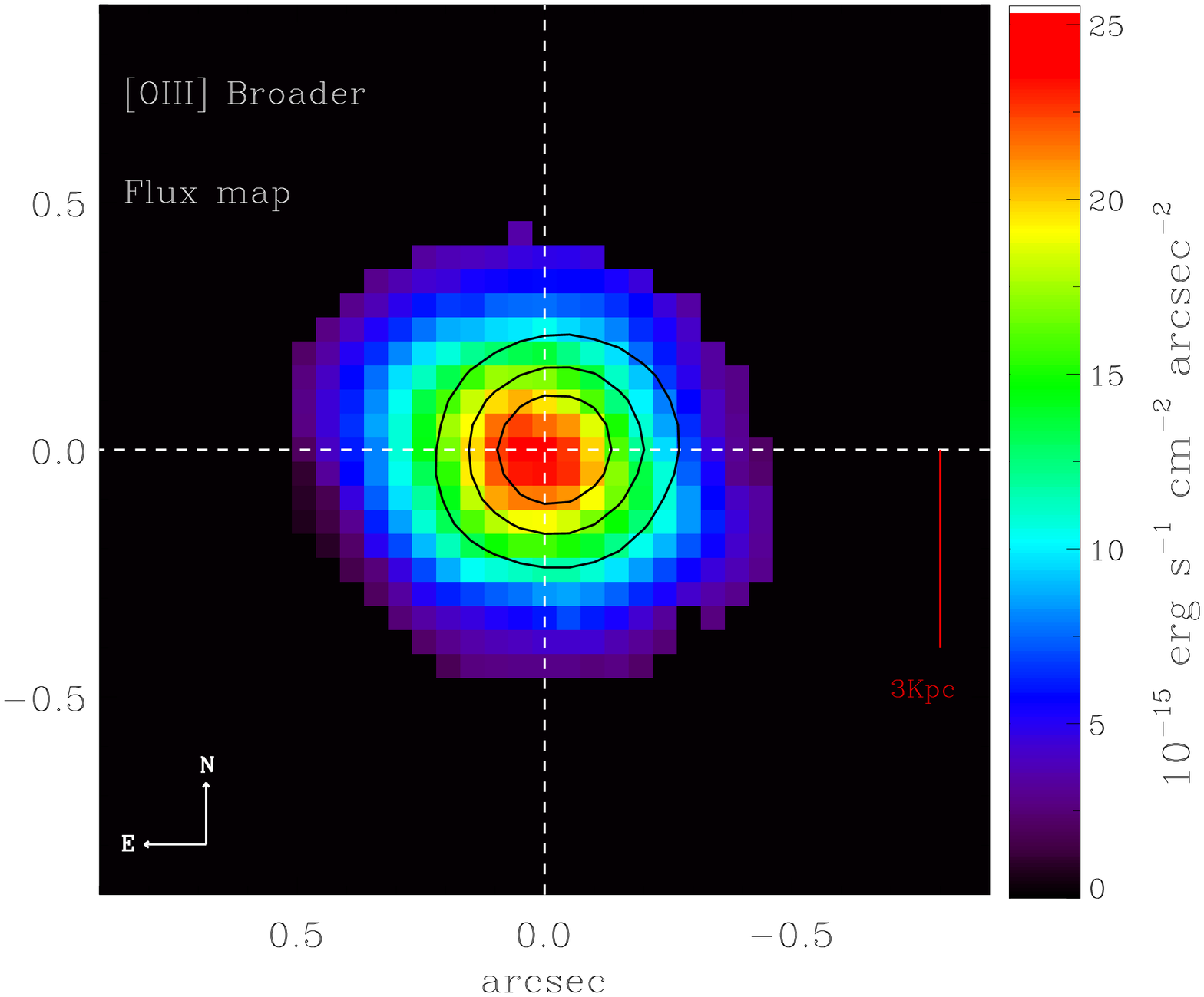}}         
  \hspace{1.7cm}
  {\includegraphics[height=0.4\textwidth,angle=90]{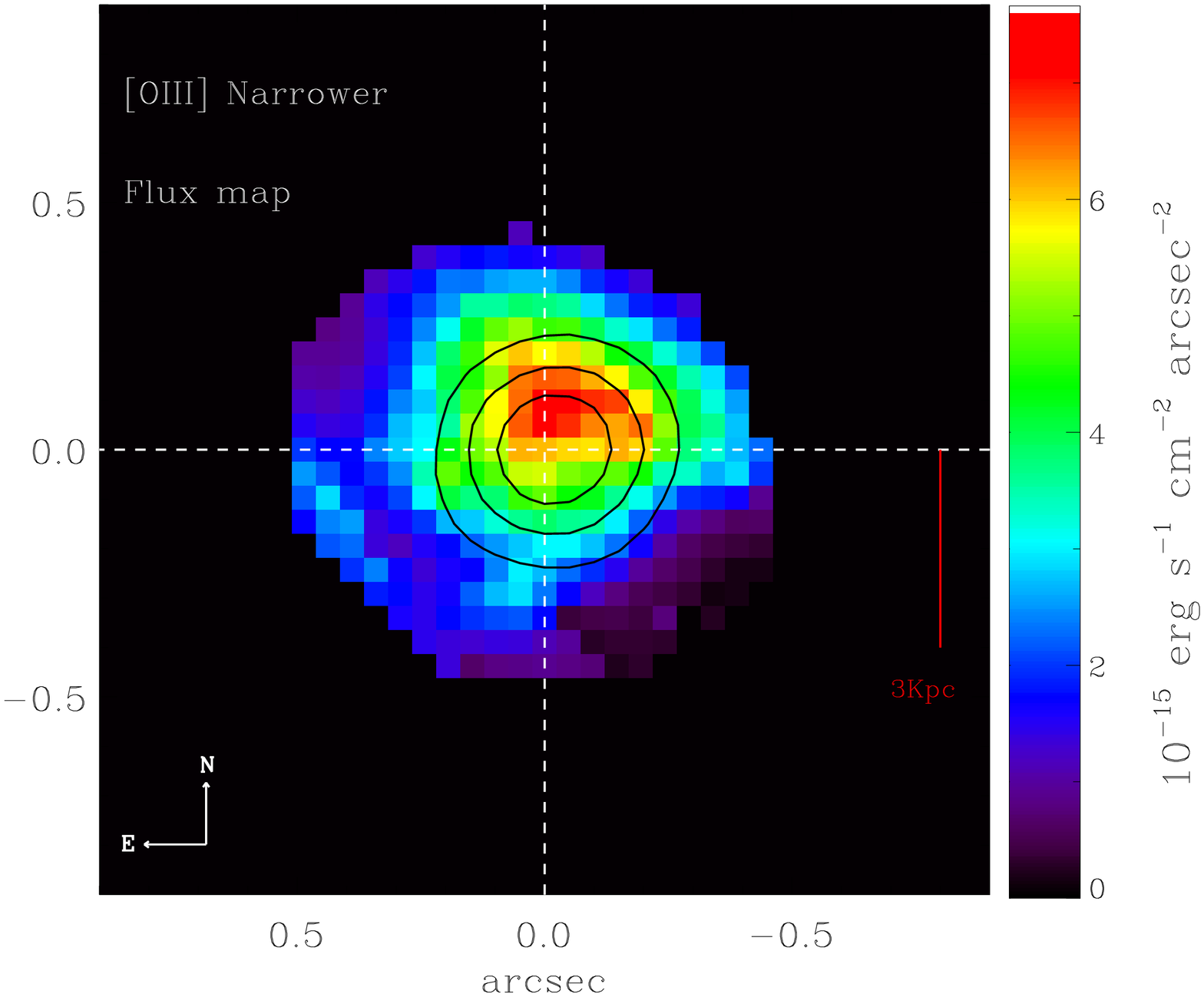}}
  \caption{Flux maps of the broad (B, left) and narrow (A, right) components of [OIII].}
  \label{fig_f_oiii}
\end{figure*}

Figure \ref{hband} shows the various components used to fit the H-band spectrum
of the central region and, in the bottom panel, the fit residuals. The region included within the
two green lines is affected by strong sky residuals and was not considered in the fit.
In this figure we can see that the [OIII] profile can be nicely fitted with two Gaussians, one
relatively narrow (A, FWHM$\sim$600~km/s) and a second one (B) blueshifted by about 700~km/s
and with FWHM$\sim$1700~km/s (much broader than typically found in the NLR
of lower luminosity AGNs). The [OIII]4959 line is fitted with the same
components, linked to have an intensity equal to one third of the [OIII]$\lambda$5007
line.
Figure \ref{fig_f_oiii} shows the flux distribution of the broad and narrow components of [OIII].
The peak of the
[OIII] broad component is shifted towards the SE, i.e. in the same direction as the strong outflow,
confirming that this is the region where the NLR develops and where the quasar radiation pressure is 
driving the outflow. The narrow component of [OIII] is instead distributed towards the N and towards
the W, i.e. similar to the H$\alpha$ narrow component, suggesting that the narrow component of [OIII]
receives a significant contribution from star formation. However, the imperfect correspondence between
the two maps suggests that a fraction of the narrow [OIII] is also contributed by the NLR.

The [OIII]4959 line is heavily blended with another line at
$\lambda _{rest}=4930 \AA$, which is also seen in the spectra of other quasars,
sometimes tentatively identified with FeII emission; however,
in the quasar discussed here (as well as in a few others showing the same feature),
the iron emission that typically encompasses the H$\beta$+[OIII] group
is particularly weak, therefore suggesting a different origin of this line.
In the residuals we observe a narrow component that is likely associated to
the same unidentified line.
The fit of the broad H$\beta$ requires three Gaussians and a powerlaw profile (as in
Nagao et al. 2006). We also included the H$\beta$ contribution associated
with the two [OIII] components. However, one should bear in mind that
the intensity of these weaker H$\beta$
components is difficult to evaluate, since these are blended within the broad, complex H$\beta$
profile.
In Table \ref{table1} we list the parameters inferred for all of
the components used in the fitting of the SINFONI spectra.

The H$\alpha$ profile (dominated by the broad component, tracing the BLR)
is clearly different from the H$\beta$ profile, which is a property that is
common to many other quasars and AGNs and that is ascribed to complex radiative transfer within the
dense gas of the BLR.
The bulk of the H$\alpha$ profile was fitted by using two very broad Gaussians (H and G in Fig.~\ref{kband}),
which give the seeing in the K-band, which is roughly consistent with the seeing measured
in the H-band observation.
As for the broad H$\beta$,
the relative intensity, shift, and width of these two lines are kept fixed over the field of view,
and their overall intensity variation reflects the seeing PSF.

We forced the inclusion of component B in the H$\alpha$ profile, by using the same velocity shift
and width as the corresponding [OIII] component, while the intensity was left free to vary.
This
component of H$\alpha$ is certainly associated with the NLR.
However, since this component is relatively broad, its intensity is poorly constrained (see uncertainty
in the flux of this component in Table~\ref{table1}), and
degenerate with the other two broad H$\alpha$ Gaussian components (as is the case for the
corresponding component in H$\beta$). The [NII] doublet associated with component B is also forced
to have the same shift and width as the corresponding [OIII] component. The relative
intensity of the two [NII]6548,6584 lines are forced to be in the ratio of 1:3.
The intensity of these [NII] lines is even less constrained than the corresponding H$\alpha$ B component, since
the wavelength of [NII]6548 nearly overlaps the intense component G of H$\alpha$.
The resulting [NII]/H$\alpha$ ratio of component B is low, but highly uncertain
($\log {(F_{[NII]}/F_{H\alpha})}=-0.73\pm0.45$), but
still consistent with the values observed in AGNs.

The narrow component ``A'' is much narrower than the other components and,
therefore, easier to disentangle from the broad H$\alpha$ profile. However, in principle, we should
include {\it two} narrow ($\rm FWHM \sim 600 km/s$)
H$\alpha$ components, one associated with the quasar NLR and another one
associated with any putative star formation in the host galaxy. However, the quality of our data
does not really allow us to fit two separate narrow components, because these would be totally degenerate.
We therefore fit a single narrow component not tied to have the same
velocity and width of component ``A'' of [OIII]. We label this narrow H$\alpha$ component
with ``A*'' (meaning that it may be partly associated with component A of [OIII], but not
necessarily). We investigate the relation of the H$\alpha$ component ``A*'' with the
[OIII] component ``A'' {\it a posteriori}. We note that it is not possible to follow a similar approach
on H$\beta$ (i.e. introduce a component ``A*'', not linked to the [OIII] components,
since, besides the problem of the blending with other components,
the signal-to-noise on H$\beta$ is much lower).

The parameters resulting from the best fit in the K-band are given in Table \ref{table1}.
We note that there is no room for narrow A* [NII]
emission, at a level of $\rm F_{[NII]6584}(A^*) < 0.21~F_{H\alpha}(A*)$. The lack of
an [NII] narrow component
is confirmed by the ``differential'' spectrum presented in Sect.~\ref{sec_ha} and in Fig.~\ref{kband}, which is
totally independent of any fitting procedure. As discussed in the body of the paper,
the lack of [NII] at a level below one fourth of H$\alpha$ indicates
that this narrow H$\alpha$ emission is mostly tracing star formation, and not the NLR.

\begin{figure*}[!]
 % \centering
  {\includegraphics[height=0.4\textwidth,angle=90]{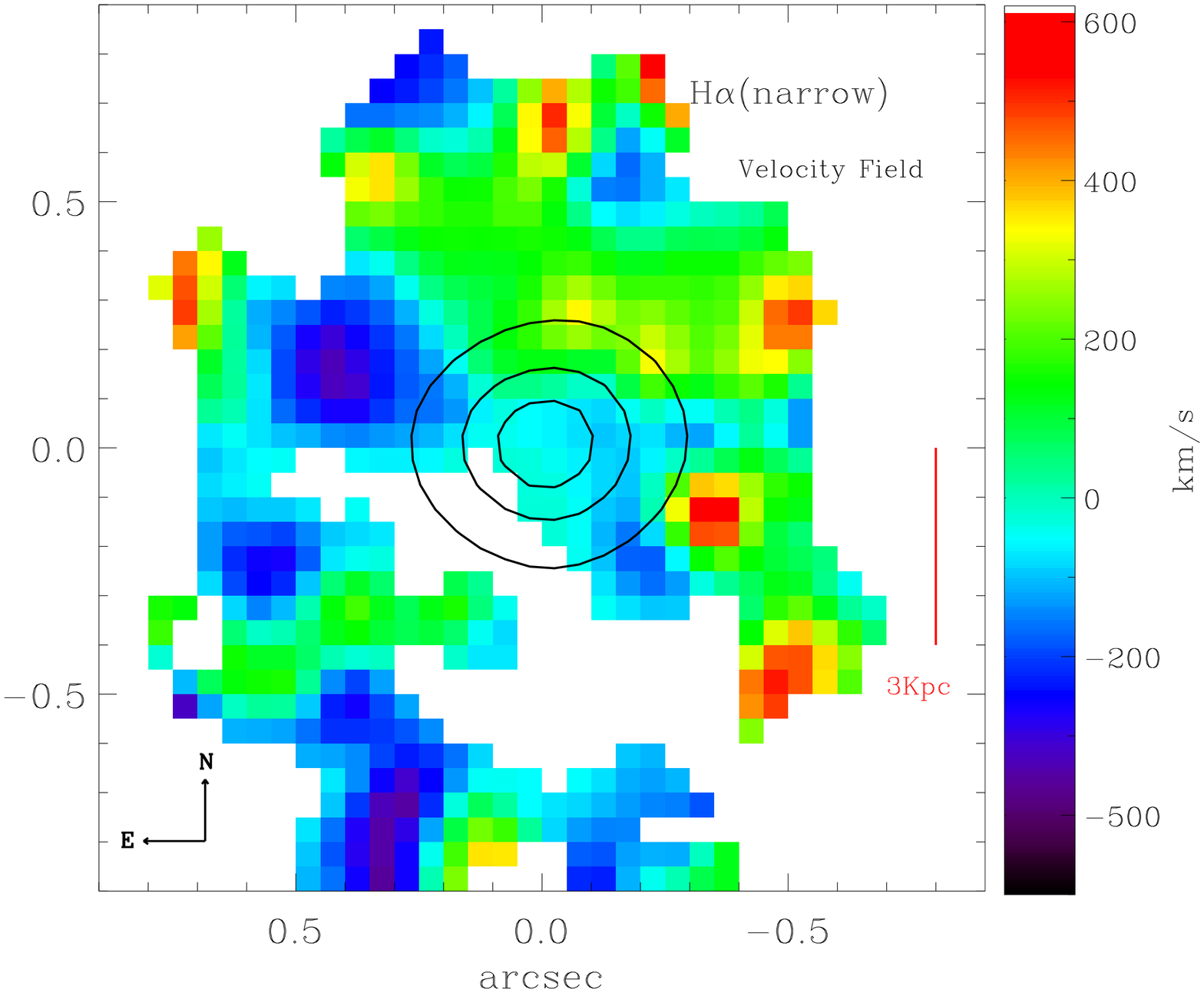}}         
  \hspace{1.7cm}
  {\includegraphics[height=0.4\textwidth,angle=90]{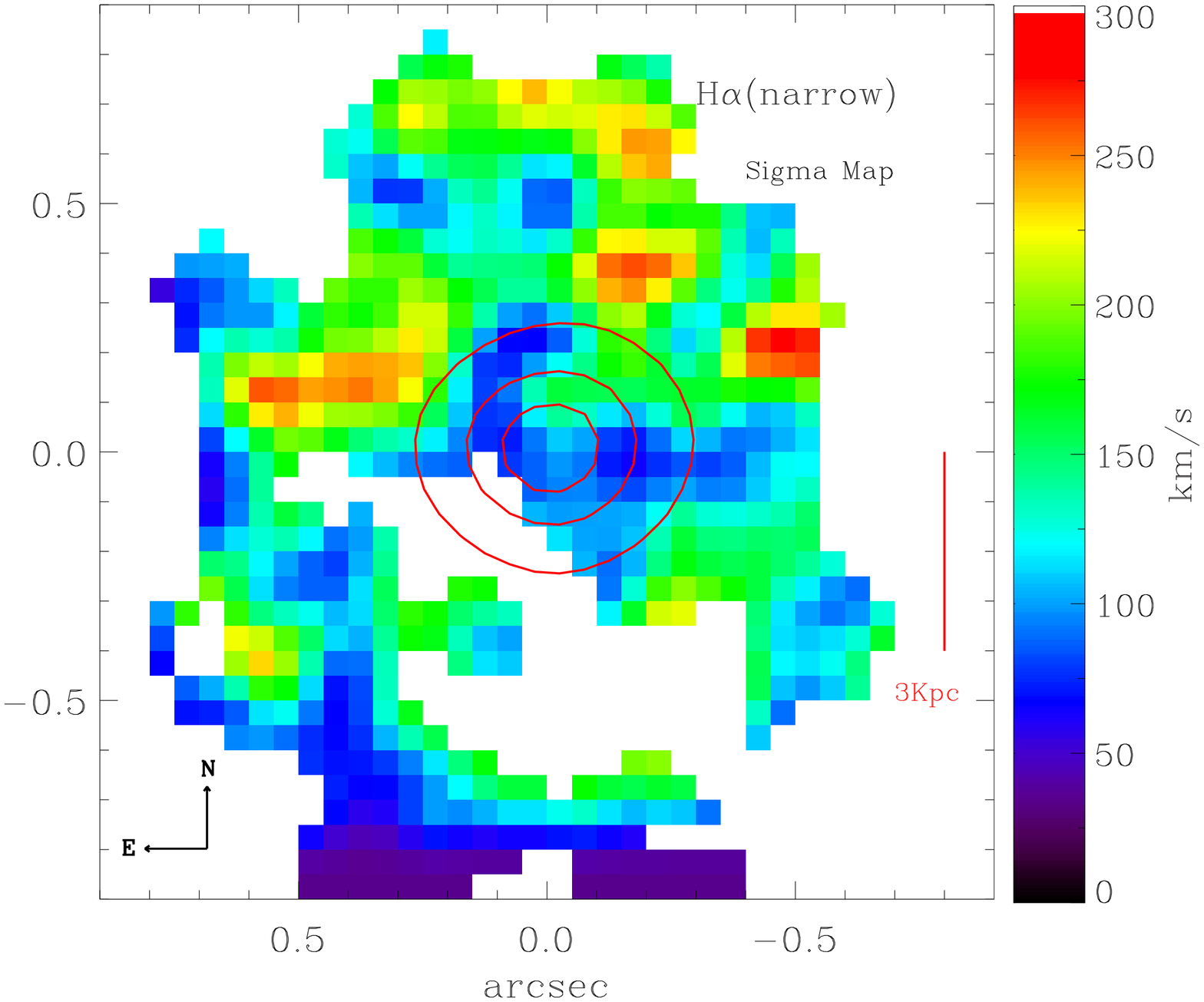}}         
  \caption{Velocity field (left) and velocity dispersion (right)
  of the narrow component (A*) of H$\alpha$.}
  \label{fig_ha_velsig}
\end{figure*}

Figure \ref{fig_ha_velsig} shows the velocity field and the velocity dispersion
of the narrow component of H$\alpha$.
Both maps are very noisy, owing to the weakness of the line. The velocity field does not clearly indicate
a rotation pattern, which would be expected by gas in a galactic disk, except possibly for an NW-SE
gradient, but the
latter may be associated with some contribution to H$\alpha$ narrow from the outflow
in the SE region.
However, only a fraction of the disk
is actually traced by the H$\alpha$ narrow,
and this, together with the
noisy velocity map, may prevent identification of
a clear rotation pattern. Moreover, it is well known that
host galaxy disks of optically selected quasars tend to be face on, as a consequence of selection effects
\citep{carilli06}, so it is not expected that quasar host galaxies have prominent rotational patterns.
The velocity dispersion map is very noisy, but it is consistent with being uniform over the area
where H$\alpha$ narrow is detected.

We finally note that the best-fit velocity and FWHM of component A* of H$\alpha$ are similar
to component A of [OIII]. This further suggests that the latter component of [OIII] is partly contributed
by the ionized gas in the star-forming regions traced by the narrow H$\alpha$. The inferred
$\rm F^{A*}_{H\alpha}/F^A_{[OIII]}\sim 0.17$, at the verge of the range typically observed in star-forming
galaxies, indicates that the flux of component A of [OIII] is not incompatible with being partly originated
by star formation, but probably a contribution by the AGN NLR is required.
As mentioned above,
the similarity of the $\rm F_{[OIII]}(A)$ map and the $\rm F_{H\alpha}(A*)$ map also supports the scenario
where part of component A of [OIII] is associated with star formation.

\begin{table*}
\begin{center}
\begin{tabular}{|l|l|l|l|l|l|l|}
 \hline
  Line & Component &  Fitting Function & $\lambda_{obs}$ & FWHM & Flux & Velocity$^1$\\
    & & & ($\mu$m) & (km/s) & ($\rm 10^{-16}erg~cm^{-2}~s^{-1}$) & (km/s) \\ \hline
  [OIII]5007      & A & Gaussian & 1.7061 &  652   &  10.9$\pm$0.56  & 92  \\
    	          & B & Gaussian & 1.7020 &  1797  & 43.6$\pm$2.3  & -663 \\ \hline
  H$\beta$        & A   & Gaussian & 1.6564 & 652  &  0.79$\pm$0.66  &  92  \\ 
    	          & B   & Gaussian & 1.6522 & 1797 &  8.8$\pm$3.6   &  -663 \\ 
	       & C & Broken Power Law & 1.6567 &  5508 & 57.9$\pm$4.6   & 156  \\ 
    	          & D & Gaussian &  1.6648 &  1150 &  4.4$\pm$2.9   & 1622  \\ 
     	          & E & Gaussian & 1.6431 &  2243  & 11.3$\pm$0.26   & -2306.6 \\ \hline 
  No ID	          & F & Gaussian & 1.6806 &  2378  & 13.18$\pm$2.3   & ---  \\ \hline
  H$\alpha$       & A* & Gaussian & 2.2361 & 616   & 1.9$\pm$0.3    & 20    \\
		          & B & Gaussian & 2.2310 & 1797   & 9.3$\pm$3.5    & -663   \\ 
	              & G & Gaussian & 2.2391 & 3386   & 73.7$\pm$5.5 & 422    \\
	              & H & Gaussian & 2.2478 & 9717   & 220.8$\pm$10.4 & 1588   \\ \hline
 [NII]6584 & A* & Gaussian & 2.2433$^1$ & 616$^1$  & $<$0.4 & 7$^2$   \\
           & B & Gaussian & 2.2382 & 1797          & 1.7$\pm$1.5 & -663   \\ \hline
\end{tabular}
\caption{Results of the spectral fit for the individual components.
\newline
The names of
the components correspond to those shown in Figs.~\ref{hband} and \ref{kband}.
Notes: $^1$ The velocities of the components are obtained by assuming
the peak of [OIII]$\lambda$5007 as a rest frame reference.
$^2$ For this [NII] line, the width and velocity were
forced, to estimate the upper limit
to the values found for the corresponding A* component in H$\alpha$.
\label{table1}}
\end{center}
\end{table*}

\section{A simple model of the ionized outflow}\label{app2} 

In this section we discuss how the physical properties of the ionized
outflow can be constrained through the observational parameters of the 
[OIII] line, by adopting a simple model for the ionized wind.
The [OIII]5007 line luminosity associated with the outflow is simply given by
\begin{equation}
\rm L([OIII]) = \int _V \epsilon _{[OIII]}~f~dV
	 \label{eq_a1}
\end{equation}
where $\rm V$ is the volume occupied by the outflowing ionized gas, $\rm f$ the filling
factor of the [OIII] emitting clouds in the outflow, and $\rm \epsilon _{[OIII]}$ the
[OIII]5007 emissivity that, at the temperature typical of the NLR ($\rm \sim 10^4~K$),
has a weak dependence on the temperature ($\rm \propto T^{0.1}$). It can be
expressed by
\begin{equation}
\rm \epsilon_{[OIII]} = 1.11~10^{-9}~h\nu _{[OIII]} ~n _{O^{+2}}~n_e ~erg~s^{-1}~cm^{-3},
	 \label{eq_a2}
\end{equation}
where $h\nu _{[OIII]}$ is the energy of the [OIII]5007 photons (in units of
$\rm erg$),
$\rm n_{O^{+2}}$ and $\rm n_e$ are the volume densities of the 
$\rm O^{+2}$ ions and of electrons, respectively (in units of $\rm cm^{-3}$).
Under the reasonable assumption 
that most of the oxygen in the ionized
outflow is in the $\rm O^{+2}$ form, then
\begin{equation}
\rm \epsilon_{[OIII]} \approx
	 5~10^{-13}~h\nu _{[OIII]}~n_e^2~10^{[O/H]} ~erg~s^{-1}~cm^{-3}
	 \label{eq_a3}
\end{equation}
where $\rm 10^{[O/H]}$ gives the
oxygen abundance in solar units.

The mass of outflowing ionized gas is given by
\begin{equation}
\rm M^{ion}_{out} = \int _V 1.27~m_H~n_e~f~dV
	 \label{eq_a4}
\end{equation}
where $\rm m_H$ is the mass of the hydrogen atom, and where we have neglected
the mass contributed by species heavier than helium.

By combining Eqs.~\ref{eq_a1} and
\ref{eq_a4} we obtain
\begin{equation}
\rm M^{ion}_{out} = 5.33~10^7~\frac{\mathcal{C}~L_{44}([OIII])}{\langle n_{e3}\rangle ~10^{[O/H]}}~M_{\odot}
	 \label{eq_a5}
\end{equation}
where $\rm L_{44}([OIII])$ is the luminosity of the [OIII]5007 line
emitted by the outflow, in units of $\rm 10^{44}~erg~s^{-1}$,
$\rm \langle n_{e3}\rangle$ ($= \int _V n_e~f~dV/\int _V f~dV$) is the average electron density
in the ionized gas clouds, in units of
$\rm 10^3~cm^{-3}$, and $\rm \mathcal{C} = \langle n_{e3}\rangle ^2 / \langle n_{e3}^2 \rangle $
is a ``condensation factor'', where $\rm \langle n_{e3}^2\rangle = \int n_e^2~f~dV/\int f~dV$.
We can assume $\rm \mathcal{C}=1$ under the simplifying hypothesis that all ionizing gas clouds have the
same density.
Also, under these assumptions, the mass of outflowing ionized gas is independent
of the filling factor of the emitting clouds.

If we assume a simplified model of the outflow (justified by the limited
information currently available to us) where the wind occurs in a conical
region, with opening angle $\Omega$, composed of ionized clouds
uniformly distributed and outflowing with velocity $\rm v$,
out to a radius $\rm R$, then the mass outflow rate of
ionized gas is given by
\begin{equation}
\rm \dot{M}^{ion}_{out} = \langle \rho \rangle _V ~v~\Omega R^2
	 \label{eq_a6}
\end{equation}
where $\rm \langle \rho \rangle _V$ is the average mass density in the whole volume occupied by
the outflow,
which is given by
\begin{equation}
\rm \langle \rho \rangle _V = \frac{M^{ion}_{out}}{V}
	 \label{eq_a7}
\end{equation}
where the volume occupied by the conical outflow is given by
$\rm V=\frac{4}{3}\pi R^3 \frac{\Omega}{4\pi}$. Unless $\rm f=1$, generally
$\rm \langle \rho \rangle  _V\neq 1.27~m_H~\langle n_e \rangle$, since the
latter numerical density (defined above) is averaged among the emitting clouds, not over the whole volume.

By replacing Eqs. \ref{eq_a5} and \ref{eq_a7} into Eq. \ref{eq_a6}
we obtain that the ionized outflow rate is given by
\begin{equation}
\rm \dot{M}^{ion}_{out} = 164~
  \frac{\mathcal{C}~L_{44}([OIII])~v_3}{\langle n_{e3}\rangle ~10^{[O/H]}~R_{kpc}}~M_{\odot}~yr^{-1}
	 \label{eq_a8}
\end{equation}
where $\rm L_{44}([OIII])$, $\rm n_{e3}$, and $\mathcal{C}$ ($\approx 1$) were
defined above, $\rm v_3$ is the
outflow velocity in units of $\rm 1000~km~s^{-1}$, and
$\rm R_{kpc}$ is the radius of the outflowing region, in units of kpc.
The outflow rate is independent of both the opening
angle $\Omega$ of the outflow and of the filling factor $\rm f$ of the
emitting clouds (under the assumption of clouds with the same density).

The kinetic power (associated with the ionized component) is then given by
\begin{equation}
\rm P^{ion}_{K} = 5.17~10^{43}
  \frac{\mathcal{C}~L_{44}([OIII])~v_3^3}{\langle n_{e3}\rangle ~10^{[O/H]}~R_{kpc}}~erg~s^{-1}.
	 \label{eq_a9}
\end{equation}

\end{appendix}

\end{document}